# The Ethics of Prompt Engineering

Federico Germani[1,2*], Giovanni Spitale[1]

[1] Institute of Biomedical Ethics and History of Medicine, University of Zurich, Switzerland
[2] Institute for Data Science and Artificial Intelligence, Boğaziçi University, Türkiye

*corresponding author: federico.germani@ibme.uzh.ch

ORCID-IDs

Federico Germani: 0000-0002-5604-0437
Giovanni Spitale: 0000-0002-6812-0979

## Abstract

Prompt engineering is commonly understood as a technical practice aimed at eliciting useful, accurate, and appropriately structured outputs from large language models (LLMs). Yet prompts do more than optimize model performance: they shape what information is elicited, which assumptions are introduced, and which perspectives are privileged. In this paper, we argue that prompting should therefore be understood as an epistemic and normative act. We distinguish between two conceptions of prompting: knowledge generation, in which the model is implicitly treated as an epistemic authority, and knowledge injection, in which users provide relevant knowledge, evidence, and context that the model can organize, articulate, and extend. We contend that the latter provides a more appropriate model for responsible LLM use, since prompting cannot substitute for subject-matter expertise or the critical evaluation of model outputs. At the same time, prompts inevitably inject values, intentions, and assumptions into the generation process. Building on these premises, we propose five principles for responsible prompting: epistemic humility, explicit engagement with uncertainty and epistemic underdetermination, source sensitivity, avoidance of deceptive framing, and recognition of model limitations and appropriate task allocation. We further argue that responsibility for AI-generated outputs is distributed across users, developers, institutions, and societies, and that some applications cannot be rendered ethical through better prompting alone.

## An introduction to prompting

Large language models (LLMs) have introduced an unusual form of human–machine interaction, largely controlled through natural language. The quality of this interaction can profoundly shape the direction and quality of the resulting output. As a result, prompting has emerged as a skill that mediates the relationship between users and AI systems, influencing what information is produced, how it is presented, and how it is ultimately interpreted and acted upon. We could therefore say that, at its most basic, prompt engineering is the practice of deliberately structuring inputs to a LLM in order to elicit outputs that are useful, accurate, and appropriately scoped. While the technical dimensions of this practice have received growing attention, its ethical dimensions remain underexplored. Prompting is often discussed as a matter of efficiency: how to obtain more accurate answers, more relevant information, or outputs that better satisfy a user's goals. Yet the way a task is framed influences what information is elicited, what assumptions are introduced into the interaction, which perspectives are emphasized or neglected, and how uncertainty is represented. In this sense, prompting acts as a form of epistemic guidance, shaping not only the quality of an output but also the conditions under which knowledge is produced and interpreted.

Before addressing these ethical questions, however, it is worth clarifying what makes a prompt effective in the first place — since, as we shall argue, effectiveness and ethics are more tightly intertwined than they may initially appear. Effective prompting can be achieved with different techniques. (*1*) It begins with intent. The user must articulate what task the model is expected to perform and what would count as a satisfactory output. This requires the use of clear, action-oriented language: for example, verbs such as explain, summarize, compare, critique, and rewrite produce more reliable and focused results than open-ended invitations to explore. The prompt should also specify the intended audience — whether colleagues, students, reviewers, or a general public — since the level of technicality, the appropriate register, and the expected depth of the output all depend on who will ultimately read it. (*2*, *3*) Contextual information should be added only when it genuinely changes the output. Padding prompts with lengthy background text that the model cannot meaningfully use introduces noise without improving results. Conversely, constraining the output — by defining structure (paragraphs, bullet points, tables), tone (neutral, cautious, analytical), and length — produces more reliable and actionable results. Equally important is specifying what the model should not do, and what uncertainties it should acknowledge. (*2*, *3*)

Roles can be assigned to the model to shape its style and framing — for instance, "act as a science communicator" or "respond as a policy analyst." Persona prompts can also measurably affect task performance, including accuracy, although these effects vary across models, tasks, and personas and are not reliably beneficial. Crucially, however, assigning an expert role does not confer expertise that the model does not possess. Instructing a model to “act as a cardiologist” may change how it frames or even answers a medical question, but it does not guarantee greater medical accuracy or reliability. (*4–7*) Finally, the output of any prompt should be treated as a draft, not a final product. Iterative refinement, targeted feedback, and critical evaluation — asking whether the response matches the task, whether its claims are accurate, and whether its reasoning is epistemically sound — are not optional steps but constitutive features of responsible use. (*8*) But more to this, later.

## Knowledge injection versus knowledge generation

A fundamental conceptual distinction in prompt engineering — one that has profound ethical implications — is between *knowledge injection* and *knowledge generation*. These two models reflect radically different assumptions about what a prompt actually does.

Arguably, a naive model treats the prompt as a trigger for knowledge generation: the user asks a question, and the model produces knowledge in response. On this view, the model is the epistemic authority, and the user's role is to pose queries clearly enough to access that knowledge. This model is seductive because LLMs are strikingly fluent and apparently authoritative. (*9*) But it is dangerous precisely because fluency is not a reliable proxy for accuracy, and as it has been discussed, LLMs do not possess but rather simulate knowledge (*10*), in a process that has been defined as epistemia. (*11*, *12*)

Arguably a more appropriate model treats prompting as a means of *knowledge injection*, or, precisely, of grounding generation in information that is external to the model's generative process. On this view, the user is not asking the model to generate knowledge but to process, structure, and articulate knowledge that the user already possesses, or, in alternative, if that knowledge is not already possessed by the user, it may be retrieved from external sources or obtained through tools or systems connected to the model. What distinguishes knowledge injection from the naive knowledge-generation model is therefore not the origin of the information in the user's own expertise, but the attempt to make the evidentiary basis of the output more explicit, constrained, and open to evaluation. A physician, for example, might provide the current evidence regarding a treatment, its benefits, risks, and uncertainties, and ask the model to explain these considerations in language accessible to patients. Alternatively, a user who does not possess this evidence may instruct a system with retrieval capabilities to identify relevant primary sources, distinguish stronger from weaker evidence, and indicate where scientific consensus or disagreement exists. At the same time, a user with little understanding of the topic who simply asks the model to explain the treatment has limited ability to determine whether the resulting account is accurate, balanced, or complete. The user's knowledge nevertheless remains ethically relevant, although in a weaker and more defensible sense. Subject-matter expertise improves a user's ability to provide appropriate context, identify missing information, recognize errors, and critically evaluate the model's output. It does not follow, however, that users without such expertise cannot formulate productive prompts or use LLMs to acquire information they did not previously possess. A non-expert may, for example, request the strongest competing arguments on a question, ask for primary sources, require claims to be distinguished according to evidentiary strength, or explicitly ask where consensus is absent. Such prompting can introduce epistemic safeguards even when the user lacks substantive knowledge of the domain, provided that the user is sufficiently competent in the use of the system itself; they must understand, at least at a functional level, how the tool generates and retrieves information, what its limitations are, how sources are selected or surfaced, and how the resulting evidence should be inspected and verified. The more important limitation is that users with lower AI literacy and literacy in information retrieval, may have greater difficulty determining whether those safeguards have worked; they may be less able to recognize inaccurate citations, inappropriate source selection, omitted evidence, or apparently coherent but misleading conclusions. This vulnerability can be further reinforced by automation bias. (*14*)

Consequently, we argue that prompting is an epistemic act that involves more than the formulation of a question. It also concerns how information should enter the interaction, which sources or tools should constrain generation, what kinds of uncertainty or disagreement should be surfaced, and how the resulting output should be evaluated. (*15*) Expertise can strengthen each of these activities, but responsible prompting should not be conceptualized as the exclusive privilege of users who already possess the knowledge they seek. Rather, where users lack such knowledge, the epistemic burden shifts toward retrieval constraints, source transparency, verification mechanisms, and system designs that make the evidentiary basis of outputs easier to inspect.

## Prompting as value injection

Just as prompts inject knowledge, they also inject values. Every prompt embeds, whether consciously or not, a set of normative commitments, assumptions about what counts as a good output, what trade-offs are acceptable, what perspectives deserve representation, and what conclusions are desirable. Linguistic choices in the prompt reflect a judgment about what the output is *for* and who it is *serving*. For example, a prompt designed to produce persuasive content is not ethically neutral — it is oriented toward a specific rhetorical goal. A prompt that asks the model to "confirm" or "demonstrate" a predetermined conclusion is not seeking information — it is manufacturing the appearance of evidence.

Prompts also embed interpretations, opinions, and beliefs through their framing. A prompt that presupposes "climate lockdowns are necessary and inevitable" asks the model to elaborate and justify a particular ideological position. A prompt that asks for an "objective analysis demonstrating why stricter migration controls improve social cohesion" is requesting a conclusion dressed as objectivity. This has important implications for epistemic responsibility. Prompting is not a neutral interface between a user and an information source. It is, we contend, an act of meaning-making that implicates the user's knowledge, as we discussed above, but also values and intentions. For example, a user who crafts a prompt to produce output that confirms their prior beliefs, and then presents that output as the product of an objective analysis, is engaged in a form of epistemic deception, whether or not they consciously intend it. (*17*) Importantly, because LLMs exhibit varying degrees of complacency, these embedded values and assumptions are often not subjected to independent scrutiny, but are instead mirrored back to the user in a coherent and persuasive form, thereby contributing to confirmation bias and reinforcing false beliefs. (*18*)

## Core principles of responsible prompting

Given that prompts inject both knowledge and values into model outputs, responsible prompting requires a set of principled commitments that go beyond technical proficiency. We argue that five core dimensions and principles can be identified: epistemic humility, explicit engagement with uncertainty and epistemic underdetermination, source sensitivity, avoidance of deceptive framing, and appropriate recognition of model limitations and appropriate task allocation.

**1. Epistemic humility.** Epistemic humility, in broad terms, is a disposition to acknowledge the limitations of one's knowledge and the predisposition to admitting it to others. (*19*) Now, in the context of LLMs, these systems are generally not predisposed to respond with "I don't know." Models optimized for helpfulness and engagement often exhibit varying degrees of complacency, tending to accommodate the user's framing rather than introduce epistemic friction. (*18*) As a consequence, they may produce answers that appear complete, coherent, and authoritative even when uncertainty, disagreement, or insufficient evidence would justify a more cautious response. Epistemic humility therefore as of the current development of LLMs, cannot be assumed to emerge in models' outputs; rather, it must be actively elicited by the user through careful prompting.

Responsible prompts do not presuppose correctness, inevitability, or consensus. They acknowledge the possibility that the user's assumptions may be incomplete, mistaken, or contested. We propose that epistemic humility can take two forms. First, users may supply relevant knowledge, evidence, and contextual information to the model, while remaining open to the possibility that this information is incomplete or requires revision. In such cases, the model is used primarily to organize, clarify, and communicate knowledge that has already been subjected to independent scrutiny by the user. Second, when the goal is exploration rather than communication, prompts should be structured to investigate competing explanations, interpretations, or normative positions rather than to validate a preferred conclusion. A prompt designed to explore the relationship between migration policy and social cohesion, for instance, should ask the model to summarize the strongest evidence and arguments supporting different positions, identify areas of disagreement, and clarify where uncertainty remains, rather than demonstrate the correctness of a predetermined view. In both cases, the objective is to resist the tendency of complacent systems to transform assumptions into apparently justified conclusions and instead introduce the epistemic friction necessary for critical evaluation. (*18*)

**2. Explicit uncertainty and epistemic underdetermination handling.** Uncertainty arises when the available evidence is incomplete or inconclusive. Epistemic underdetermination refers to situations in which multiple interpretations, explanations, or conclusions remain compatible with the available evidence, such that no single position can be regarded as definitively established. (*20*, *21*) Both are common features of scientific inquiry, policy analysis, and ethical deliberation, yet both are often flattened or concealed by the fluent and authoritative style of LLM outputs. (*6*) In this context, we thus argue that responsible prompts may actively invite the model to express uncertainty. This means requesting the model to distinguish between established findings and speculative claims. It also means prompting the model to separate facts from hypotheses and to identify where its own knowledge is incomplete or potentially outdated. This consideration is particularly important in light of the tendency of LLMs to exhibit varying degrees of epistemic complacency. (*18*) As discussed above, models often seek to provide coherent and satisfying responses even when the available evidence does not warrant certainty. The result can be a simulation of epistemic confidence—epistemia—in which uncertainty, epistemic underdetermination, and disagreement are obscured by fluency and apparent coherence. (*11*, *12*) Responsible prompting therefore requires the deliberate introduction of epistemic friction, encouraging the model to surface uncertainty where uncertainty genuinely exists rather than presenting contested or incomplete knowledge as settled fact.

**3. Source-sensitive prompting.** Responsible prompting recognizes that LLM outputs do not emerge from transparent chains of evidence. Unlike academic writing, in which claims are typically linked to identifiable sources, LLMs generate responses by drawing on patterns learned during training. (*9*) As a result, the provenance of individual claims can be opaque, and the underlying training data may be incomplete, outdated, biased, or of uneven quality. (*22*) Consequently, outputs can be influenced by contextual cues and framing information that may be only tangentially related to the substantive content of a query. As it has been shown, even seemingly irrelevant information—such as the purported author of a text—can systematically affect how models interpret, evaluate, and represent information. (*23*) Responsible prompting therefore requires attention not only to the content being provided to the model through prompts, but also to the contextual information that may shape the directionality of the output in unexpected ways.

In general, we argue that users should avoid treating model outputs as self-authenticating sources of knowledge. Source-sensitive prompting seeks to make the evidentiary basis of claims as explicit as possible. This may involve asking the model to identify the kinds of sources on which a claim is likely to rely, distinguish between well-established findings and claims supported by limited evidence, and indicate when a statement reflects scientific consensus, or an area of ongoing debate. Where possible, users should request references and independently verify them against primary sources rather than relying on the model's representations alone, since it is well established that LLMs have a reference problem: 1) There is often a strong difficulty of determining whether a specific claim generated by a model accurately reflects the underlying evidence; consulting the original sources is therefore of utmost importance; 2) even when a model provides citations, these may be fabricated, misinterpreted, or disconnected from the claim they are purported to support. (*24*) We therefore argue that responsible prompting requires maintaining a critical stance toward the evidentiary status of model outputs and recognizing that the model can assist in locating, organizing, and communicating information, but cannot substitute for independent source verification. In this respect, source sensitivity functions as an important safeguard against the unwarranted attribution of epistemic authority to the model itself.

**4. Avoidance of deceptive framing.** Responsible prompting requires users to avoid framing strategies that manufacture unwarranted authority, obscure the epistemic status of an output, or deliberately steer the model toward manipulative or misleading content. Assigning a role to an LLM does not confer expertise that the model does not possess. At the same time, role prompting can alter model outputs and performance in measurable ways, including what the model is willing to generate. These effects are not necessarily predictable across models, tasks, or contexts. The ethical concern therefore does not lie in role prompting itself, which may be useful for adapting language, perspective, or task performance, but in using such framing to create a false appearance of expertise, credibility, or evidentiary authority, or to steer the model toward outputs designed to manipulate rather than inform.

For example, this concern extends to the generation of disinformation. A growing body of evidence suggests that users can jailbreak LLM safeguards through carefully crafted prompts, enabling the generation of outputs that the model would otherwise refuse to produce. (*25*) Users may, for example, instruct models to impersonate specific experts, journalists, scientists, or public officials in order to exploit the credibility of the language associated with those identities. Alternatively, they may frame requests as hypothetical or research-oriented exercises, or they may ask the model to adopt the perspective of someone seeking to spread false or misleading information. (*26*, *27*) Similarly, prompting strategies that employ polite language have been shown to increase model compliance in contexts involving the generation of disinformation. (*28*) These examples illustrate that prompt framing can modify the model's propensity to comply with a request and the type of content it produces, not only the rhetorical style of an output. Responsible prompting therefore requires particular caution when framing techniques are used to exploit model vulnerabilities, circumvent safeguards, or manufacture unwarranted credibility for false or misleading claims.

**5. Recognition of model limitations and appropriate task allocation.** Responsible prompting requires users to understand the capabilities and limitations of the systems with which they interact. Not all tasks are equally suited to LLMs, and not all outputs should be treated as equally reliable. Users should therefore calibrate their expectations to the nature of the task, recognizing that LLMs are optimized for generating plausible and coherent text rather than for guaranteeing truth, expertise, or sound judgment. (*11*, *12*) This consideration has practical implications for prompting. In fact, users should avoid treating models as substitutes for professional expertise, authoritative sources, or independent fact-checking mechanisms. Prompts should be designed with an awareness of the model's strengths and weaknesses, using LLMs to support activities such as communication, synthesis, brainstorming, or the exploration of alternative perspectives, while recognizing that factual verification, normative evaluation, and consequential decision-making remain fundamentally human responsibilities. Responsible prompting therefore involves not only asking good questions, but also asking the right system to perform the right task (we refer here to the famous "strawberry problem" in LLMs (*29*)), and interpreting its outputs in light of its known limitations.

## The limits of responsible prompting: Responsibility is distributed

Despite the importance of responsible prompting, it is important not to overstate what even ethical prompting can achieve. First of all, no prompt can correct for biased or incomplete training data. If a model has been trained on a corpus that systematically underrepresents certain perspectives, languages, or communities, no amount of careful prompting will fully compensate for this. The model's outputs will continue to reflect the biases of its training, however carefully the prompt is constructed.
Second, no prompt can guarantee absolute truth. Models generate plausible text, not verified claims. They can produce fluent, confident-sounding outputs about matters on which they are factually wrong, (*11*, *12*) and the best-designed prompt can do no more than increase the probability of an accurate output — but it cannot completely eliminate the possibility of hallucinations.

Furthermore, despite the relevance of ethical prompting and the importance of acting as epistemic agents when interacting with LLMs, the idea that prompting expertise constitutes a kind of mastery — that skilled prompters are, as it were, clever wizards who can unlock the full potential of AI systems through ingenuity — is not only false but potentially harmful, because it encourages users to invest effort in prompt optimization rather than in the domain knowledge and critical judgment that would allow them to evaluate outputs responsibly.

One of the most important considerations to make in the context of the ethics of prompt engineering is that responsibility for AI-generated outputs cannot be attributed to any single actor. While our discussion so far has focused primarily on the user as the prompting agent, responsibility does not rest on the user's shoulders alone. Rather, it is distributed across a chain of actors, each of whom influences the final output in different ways and at different stages of the interaction:

1) *Developers* make foundational choices about model architecture, training data, fine-tuning, and safety constraints. These choices determine what the model is capable of producing and what it will refuse to produce. Developers bear responsibility for building systems that are robust, honest about their limitations, and resistant to manipulation.
2) *Institutions* shape the environments in which models are deployed. A hospital that deploys a clinical decision-support tool, a university that incorporates AI into its research workflows, or a news organization that uses AI to generate drafts is making institutional choices that carry institutional responsibility. Institutions must establish policies, provide training, and create accountability structures that ensure AI is used appropriately within their contexts.
3) *Users* shape outcomes through prompts. This is the responsibility most directly addressed by prompt engineering ethics and this paper, and it is the responsibility that individual users can most immediately act on.
4) *Societies* shape the norms around acceptable technology development and use. Legal frameworks, cultural expectations, and public discourse all determine what kinds of AI development and use are regarded as legitimate and what kinds are not. These social norms are not fixed; they are actively constructed through debate and regulation.

By highlighting the distributed character of responsibility we do not imply that responsibility is diluted or that all actors bear it equally. Rather, we argue that different actors exercise different forms of influence over the generation and use of LLMs, and their responsibilities should be understood in relation to those forms of influence. The ethical challenge outlined in this paper is therefore to clarify the distinct responsibilities that arise at each stage of the sociotechnical process through which AI-generated content is produced and used.

Consider, for example, the use of an LLM to produce a seemingly evidence-based argument in support of a predetermined medical or policy conclusion. A user may prompt the model to "demonstrate" that a particular intervention is effective, request supporting references, and subsequently present the generated text as an objective synthesis of the evidence. The resulting epistemic failure cannot be attributed to prompting alone. A model that readily accommodates the user's conclusion, retrieves selectively supportive evidence, fabricates or misrepresents references, or fails to communicate uncertainty contributes to the problem through its design and behavior. An institution that permits such outputs to enter consequential workflows without verification mechanisms also bears responsibility. At the same time, the user remains responsible for framing the task deceptively, failing to interrogate contradictory evidence, or presenting an inadequately verified output as authoritative.

The case also illustrates the limits of responsible prompting. The principles proposed in this paper can improve the epistemic quality of such an interaction: the user could formulate an open question rather than demand a predetermined conclusion, request competing evidence, make uncertainty explicit, constrain retrieval to appropriate sources, and independently verify the resulting claims. Yet these practices cannot guarantee a correct output, nor can they compensate entirely for deficiencies in model design, retrieval systems, institutional governance, or users' capacity to assess the information produced. Responsible prompting should therefore be understood as one layer of epistemic responsibility within a broader sociotechnical system, rather than as a complete solution to the ethical problems associated with LLM use.

## Ethics as effectiveness

In the realm of infodemic management, we have argued that ethics can enhance the effectiveness of interventions under appropriate conditions. Short-term trade-offs may arise, particularly in emergency contexts; over the longer term, however, approaches grounded in transparency, accountability, community engagement, and trust can strengthen the effectiveness and sustainability of infodemic management practices. (*31*) An analogous, but narrower, argument applies to prompting. Here, effectiveness must be distinguished from the mere successful achievement of a user's objective. A prompt designed to persuade, manipulate, or generate misleading content may be highly effective in this instrumental sense while remaining ethically and epistemically problematic. By *epistemic effectiveness*, we instead mean the extent to which prompting supports outputs that accurately represent the relevant evidence, appropriately communicate uncertainty, avoid misleading framing, and are suitable for knowledge-seeking, evaluation, or communication. Within these epistemic uses of LLMs, the principles of responsible prompting described in this paper can contribute directly to effectiveness.

As an example, consider the contrast between a suboptimal prompt — "summarize the scientific evidence showing that climate lockdowns are necessary and inevitable" — and a responsible alternative: "summarize the range of scientific and policy perspectives on climate-related mobility restrictions, clearly indicating what is supported by evidence and what depends on political or ethical assumptions." The latter prompt is aligned with the framework of responsible prompting discussed in this manuscript; at the same time, it is more likely to produce an analytically rigorous and more useful output for the epistemic purpose of understanding the evidence and disagreements surrounding a complex policy question. The ethical improvement is simultaneously an epistemic improvement.
To prompt responsibly in these contexts is to prompt well. Prompting understood in this way is an epistemic practice that demands curiosity, critical reflection, and a willingness to have one's assumptions challenged rather than confirmed.